\documentclass[structabstract]{aa}  
\usepackage{graphicx}
\usepackage{txfonts}
\usepackage{natbib}
\usepackage{txfonts}
\usepackage{fixltx2e}
\bibpunct{(}{)}{;}{a}{}{,}
\def\g{\hbox{$\gamma$}}

\def\psrb{PSR\,B1259$-$63}
\def\psrbls{PSR\,B1259$-$63/LS2883}
\def\1303{HESS\,J1303$-$631}
\def\lsi{LSI+61\,303}
\def\ls{LS\,5039}
\def\hess{H.E.S.S.}
\begin{document}
   \title{TeV Flux modulation in \psrbls}

   \author{M.~Kerschhaggl\thanks{now at Physikalisches Institut, University of Bonn, Nussallee 12, Bonn, 53115, Germany \email{mkersch@physik.uni-bonn.de}}
          }

   \institute{Department of Physics, Humboldt University, Newtonstrasse 15, Berlin, 12489, Germany\\
              \email{mkersch@physik.hu-berlin.de}
             }

    \date{Preprint online version: September 27, 2010}
%
%
 
  \abstract
   {\psrbls~ is a binary system where a 48 ms pulsar orbits a massive Be star with a highly eccentric orbit ($e=0.87$) with a period of 3.4 years. The system exhibits variable, non-thermal radiation visible from radio to very high energies (VHE) around periastron passage. This radiation is thought to originate from particles accelerated in the shock region between the pulsar wind (PW) and stellar outflows.}
   {The consistency of the \hess~ data with the inverse Compton (IC) scenario is studied in the context of dominant orbital phase dependent adiabatic losses.}
   {The dependence of the observed TeV flux with the separation distance is analyzed. Model calculations based on IC scattering of shock accelerated PW electrons and UV photons are performed. Different non-radiative cooling profiles are suggested for the primary particle population to account for the variable TeV flux.}
   {The TeV fluxes obtained with \hess~in the years 2004 and 2007 seem to be only dependent on the binary separation. The presented results hint at a peculiar non-radiative cooling profile around periastron dominating the VHE emission in \psrb. The location of the stellar disc derived from this non-radiative cooling profile is in good agreement with that inferred from radio observations.}
{}


   \maketitle

\section{Introduction}
The binary system \psrbls~comprises a pulsar with a period of 48 ms and a massive star of spectral type B2Ve featuring a wind with a polar and equatorial component \citep{john92}. The latter slower and denser wind constitutes a circumstellar disc. The interaction of the pulsar with the dense environment (radiation and/or matter) in the binary system may give rise to non-thermal processes. \cite{tavani&arons} have discussed various plausible scenarios for the production of non-thermal emission in the system. 
The following will focus on the non-thermal processes that occur in the shocked pulsar wind (PW)  \citep{kennel_coroniti}.
Indeed, the supersonic PW is terminated at a strong shock wave, and at this shock particles may be accelerated up to very high energies (VHE).  These accelerated particles could, in case of a leptonic PW, upscatter photons from the companion star to VHE via inverse Compton (IC) processes producing VHE emission \citep{Tavani94}.
In the framework of this scenario, \citet{kirk} have predicted the level and orbital phase dependence of the $\g$-ray flux. Two regimes of non-thermal particle cooling were considered: (i) negligible non-radiative ({\it adiabatic}) energy losses (i.e.~dominance of radiative losses); and (ii) dominant (orbital phase independent) adiabatic losses.
However, the observations  of \psrb~ with the \hess~ telescope array \citep{1259} found  a lightcurve that  significantly deviates from the two lightcurves obtained by \citet{kirk}.
This discrepancy supported hadronic scenarios for the production of  VHE emission in the system. In the framework of this scenario, VHE protons originating from the pulsar interact with the dense outflow from the companion star, producing VHE emission. A characteristic feature of this scenario are two sharp maxima in the VHE emission related to the pulsar passage through the dense equatorial outflow of the star \citep{kawachi}. The importance of the dense circumstellar disc seems also to be indicated by the lack of pulsed radio emission shortly before and after periastron \citep{J05}.
Moreover, correlations between the X-ray and TeV fluxes seemed to support the hadronic scenario for the VHE radiation \citep{XMM}, since in the case of proton-proton interactions not only VHE \g-rays are produced, but also secondary leptons. These secondary particles may effectively emit through the IC and synchrotron channels providing the X-ray and radio emission correlated with the VHE energy band \citep{XMM}. Thus, the TeV flux from \psrb~ should be related to the rate of primary interactions, i.e.~to the location of the circumstellar disc on the pulsar orbit. However, it should be noted that the reconstruction of the disc properties from the eclipse of radio pulses leads to somewhat different values for the geometrical parameters of the stellar disc than those required in the hadronic scenario  \citep{Bmz05}. Moreover, \hess~data taken around the 2007 periastron covering orbital phases different to the ones  in 2004 seem to contradict such a hadronic disc scenario since the TeV flux does not follow a simple disc density profile as indicated by 2004 data  \citep{1259new}. 
Thus, the currently obtained lightcurves apparently do neither support leptonic nor hadronic scenarios for the production of VHE photons in \psrb. This discrepancy suggests the existence of additional effects which have to be considered in the modelling.  In the case of the hadronic scenario, the simplest assumption is the orbit-to-orbit variability of the disc location. This hypothesis may be supported by the orbit-to-orbit variability of the pulsed radio emission detected from the system \citep{wang} but seems unlikely since there are no significant inter-orbital variations of X-ray and unpulsed radio fluxes \citep{suzaku, J05}. Clarifying this contradictory behavior with respect to a hadronic disc scenario requires a dedicated study. In what follows, a possible origin of the complicated behavior of VHE fluxes in the framework of a leptonic scenario will be discussed.\\ %
The IC TeV energy flux is expected either to peak at periastron, where the target photon field density is maximal for the standoff point, i.e.~at the location of the pulsar-star wind shock, or to show a rather smooth orbital phase dependence in case of a saturation regime \citep{kirk}. This is, however, in opposition to what was measured by \hess~  The measured VHE lightcurve clearly shows a drop in photon flux towards periastron. Moreover, the orbital phase dependence of the X-ray lightcurve \citep{XMM} does not fit the shape expected in a simple leptonic model \citep{kirk}. All these discrepancies ask for a more sophisticated modelling. \citet{khangulyan} modelled  the VHE data obtained with \hess~in 2004 with synchrotron/IC scenarios that were accounting for (i) a change of the acceleration rate due to enhanced losses close to the periastron passage, (ii) dominant orbital phase dependent adiabatic losses, (iii) the orbital phase dependent change of the pulsar wind bulk Lorentz factor due to the interaction with stellar photons. More recently, \citet{uchiyama} modelled the X-ray data obtained with the Suzaku X-ray satellite with a synchrotron and IC radiation model that included orbital phase dependent adiabatic losses. Finally, \citet{takata} suggested that the observed variability of non-thermal fluxes is related to the orbital phase dependence of (i) the pulsar wind bulk Lorentz factor and magnetization parameter, (ii) the magnetization parameter and the spectral power law index of shock-accelerated particles. In  this last approach, non-radiative losses were not accounted for.\\  
The introduction of orbital phase dependent non-radiative losses accounts for the following physical effects most likely being present in \psrb: (i) the post-shock flow propagating in the confined region formed by the PW and the shocked stellar wind \citep{bogovalov} should suffer significant adiabatic losses \citep{khangulyan_hepro}; (ii) the loss rate depends on the hydrodynamics of the interaction and may thus naturally depend on the orbital separation distance and the density of the stellar wind. Moreover, since the pulsar wind is expected to be anisotropic \citep{bogo&khang}, the adiabatic loss rate could  have a rather complicated orbital phase dependence. 
Hence, in the following the approach of \citet{khangulyan} is used, taking into consideration the data obtained with \hess~in 2007.\\ 
This paper is organized as follows. First of all, a phenomenological discussion of the \psrb~lightcurve both in the \g-ray and X-ray bands is presented in Sec.~\ref{lc}.  A comparison of the expected IC TeV energy flux with the \hess~data allows an inference of the non-radiative cooling profile. This ansatz yields results for the prediction of TeV photons from \psrb~that describe  the peculiar drop at periastron with reasonable accuracy \citep{khangulyan}. In their \citeyear{khangulyan} paper, \citet{khangulyan} used the \hess~2004 data in order to calculate the non-radiative cooling coefficients for the \psrb~orbit. In Sec.~\ref{model} of this article their analysis is redone with the addition of VHE data from the 2007 periastron passage. Moreover, three different possible cooling profiles are compared.
\section{The lightcurve of \psrb~as seen in X-rays and \g-rays}          
\label{lc}
\begin{figure}[h!]
\centering
\resizebox{\hsize}{!}{\includegraphics{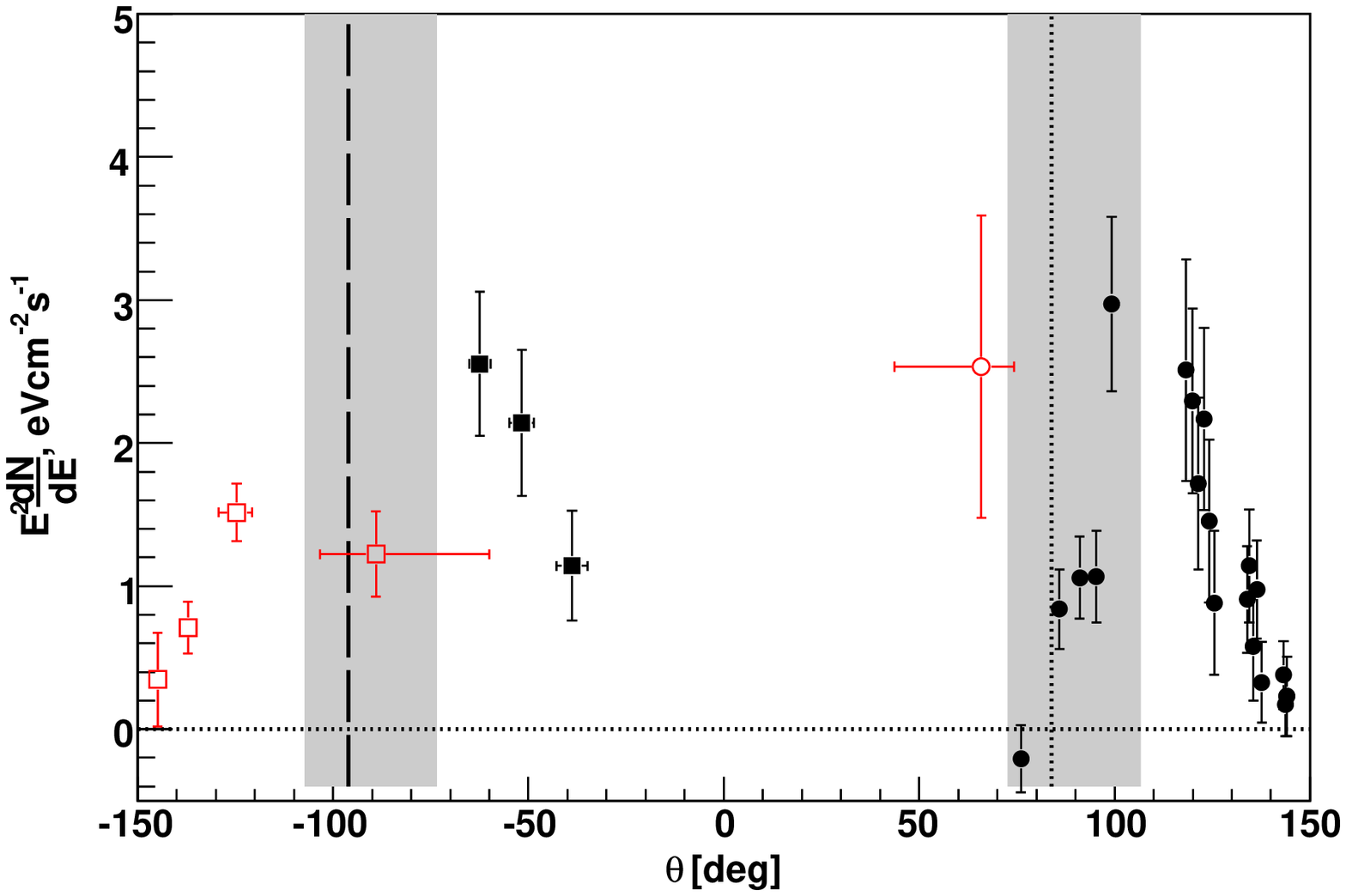}} %
\resizebox{\hsize}{!}{\includegraphics{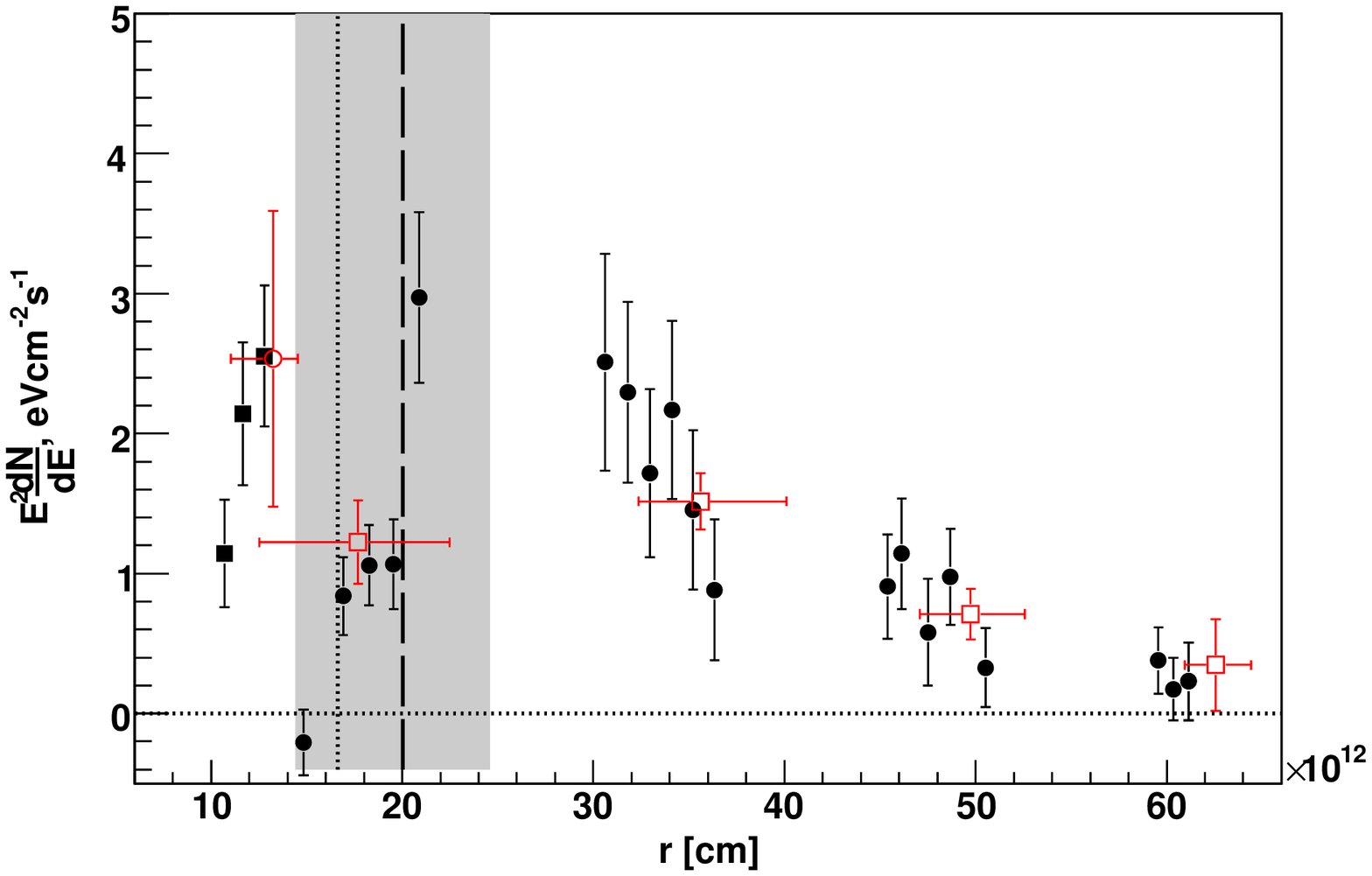}} %
\caption{Energy fluxes of 1 TeV photons from \psrb~as a function of the true anomaly (top panel) and of the binary separation distance (bottom panel). The measurements were carried out by \hess~during the 2004 (black full  symbols  - fluxes using a bin width of 2 days) and 2007 (red empty  symbols  - monthly fluxes) periastron passage, respectively. Squares denote pre- and circles post periastron measurements, respectively. The vertical grey boxes indicate the location of the stellar disc inferred from the best fit geometrical parameters from radio data. The vertical dashed and dotted lines denote the positions of the stellar equatorial plane along the pre- and post periastron orbit, respectively \citep{Bmz05}.}
\label{curve}
\end{figure}

While \g-ray binaries, such as \ls~and \lsi, have been observed over several orbits and thus could be confirmed as periodical VHE emitters \citep{5039,LSI2} this was not possible for \psrb~so far because of the rather long orbital period of 3.4 years. The overlap between the \hess~datasets of 2004 and 2007 is marginal and thus inconclusive in this regard (see top panel in Fig.~\ref{curve}). It is thus up to now unknown whether the TeV lightcurve of this object shows periodicity. There is, however, some strong hint  that the object indeed behaves periodically  in the TeV regime: the lower panel in Fig.~\ref{curve} shows the energy flux of 1 TeV photons of \psrb~as a function of the separation distance $r$ between the pulsar and its companion star. The bidaily 2004 energy fluxes for 1 TeV photons are taken from \citet{1259} (see their Fig.~8) whereas the 2007 fluxes are calculated from bolometric fluxes above 1 TeV assuming the underlying spectral power law as described in \citet{1259new}. 
There is an apparent relation  between pre and post periastron data obtained  in the years 2004 and 2007 by \hess~ and  the orbital distance. The correlation coefficient between the 2007 fluxes shown in time bins of months and the monthly weighted mean of the bidaily 2004 data as a function of $r$ amounts to $\rho=0.89\pm0.12$. Despite the limited statistics of the TeV data and the fact that firm conclusions can not be drawn at present, this behavior would indicate essentially two things:  (i) the TeV flux is symmetric with respect to periastron and thus mainly a function of the binary separation, (ii) the system shows indirect signs of periodicity. It should be noted that in frameworks of a leptonic scenario this kind of behavior is rather unexpected. Indeed, since any two different orbital positions, which are symmetric with respect to periastron, are characterized by different IC scattering angles (according to the currently accepted orbital elements), one should expect different fluxes in the direction of the observer due to anisotropic IC scattering \citep{kirk, aharonian}. The significance of this effect depends on many different factors: the actual location of the production region, the $\gamma$-ray energy band, the slope of the electron spectrum and the temperature of the optical star. Another source of asymmetry in \psrb~should be Doppler modulation of the post shock flow of relativistic particles. The orbital parameters of the system in connection with expected large Lorentz factors of the bulk flow imply a strong phase dependence of the related non-thermal emission \citep{khangulyan_hepro}. Taking into account the system orientation with respect to the observer \citep{john92}, this naturally should lead to an asymmetry in the observed VHE fluxes.  
\begin{figure}
\centering
\resizebox{\hsize}{!}{\includegraphics{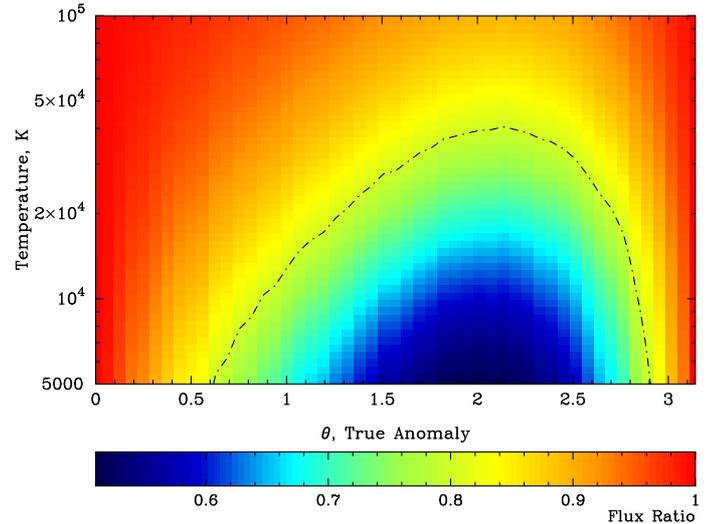}} 
\caption{Ratio of VHE fluxes (integrated  between 380GeV--10TeV) for any two  different points of the orbit characterized by the same distance to the optical star as a function of star temperature and orbital phase. The distribution of scattering  electrons was assumed to have a power law index equal to $2$. The dashed-dotted line corresponds to the contour line for a flux ratio of $0.8$. Given the relatively large statistical and systematic uncertainties (close to $20\%$) of the VHE data, in case of a star temperature close to (or higher than) $\sim4\times10^4$~K, the TeV light curve might look symmetric with respect to the periastron passage. However, if the star temperature is lower (e.g. $\sim2\times10^4$~K), the difference in the fluxes for orbital locations with a true anomaly of $\theta=\pm 115^\circ$ (or $\theta=\pm2$~rad) should be measurable with H.E.S.S.}
\label{fig_anisotropy}
\end{figure}
\begin{figure}%
\centering
\resizebox{\hsize}{!}{\includegraphics{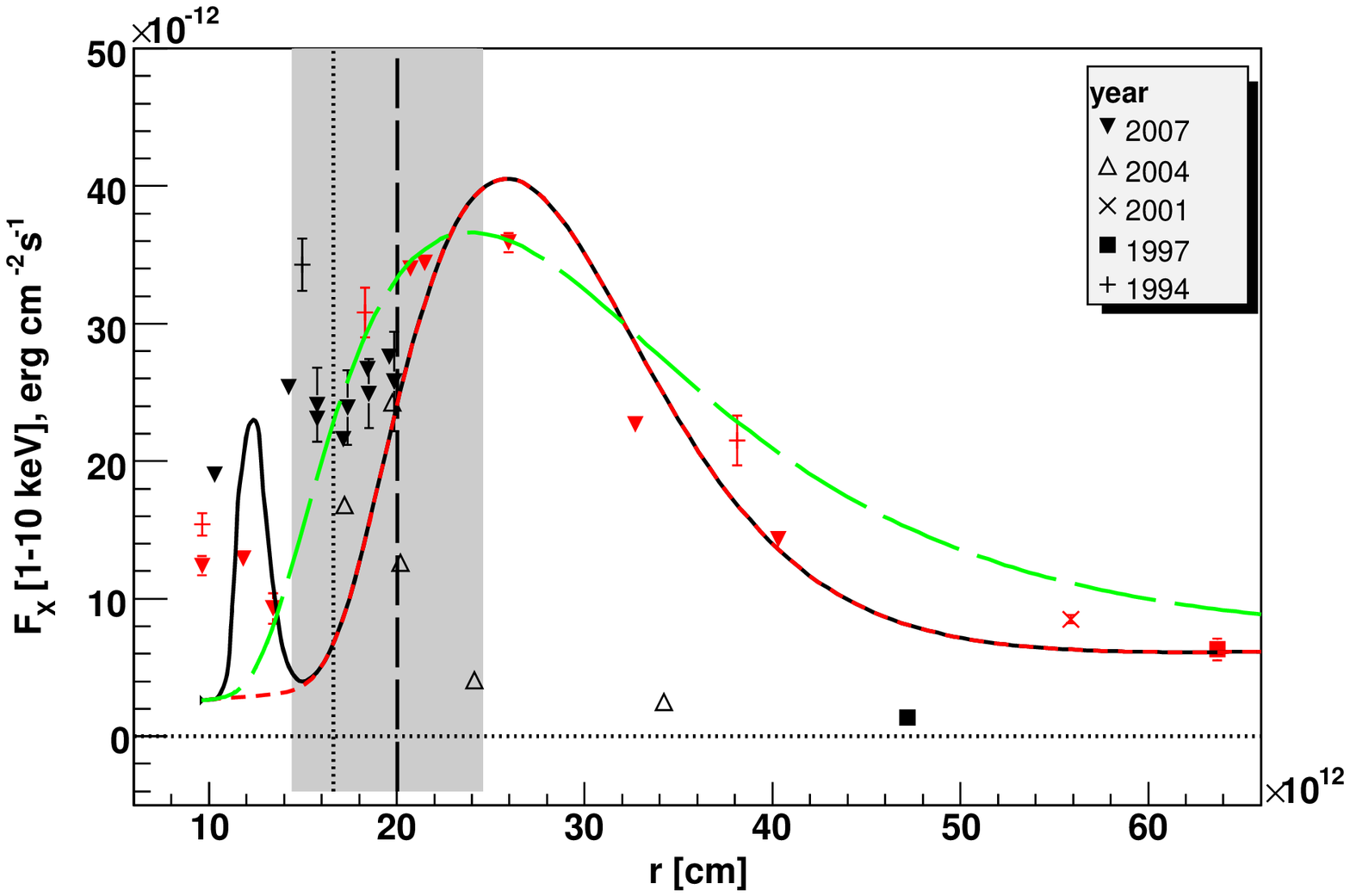}} 
\caption{X-ray fluxes in the 1$-$10 keV band from pre periastron (black symbols ) and post periastron (red symbols ) phases measured in the years 1994, 1997, 2001,  2004 and 2007 as a function of the binary separation distance $r$. Also shown are three different model X-ray light curves corresponding to the adopted cooling profiles shown in Fig.~\ref{cool}. The grey band indicates the extension of the stellar disc. The vertical dashed and dotted lines correspond to the location of the stellar equatorial plane along the pre- and post periastron orbit, respectively \citep{Bmz05}.}
\label{x-ray}
\end{figure}
In Fig.\ref{fig_anisotropy} the ratios of the VHE fluxes (integrated  between 380 GeV and 10 TeV) for the locations of the pulsar symmetric with respect to periastron passage (i.e.~characterized by the same absolute value of the true anomaly) are shown. This figure illustrates the dependence of the flux ratio not only on the pulsar orbital location, but also on the star temperature. For the calculations the electron distribution was assumed to be a power law with index $2$. For the currently accepted star temperature $T_*=2.3\times10^4\rm K$, the fluxes should differ by a factor of $1.5$ at the locations with true anomaly $\theta\sim2\rm rad$ (or $\theta\sim115^\circ$, $r\sim3\times10^{13}\rm cm$). Apparently, the \hess~observations from these epochs show almost equal fluxes (see Fig.\ref{curve}) within the limited statistics of the data. In case this feature turns out to be significant  there are several possible explanations in the IC scenario for such an  observational result. 

(i) The ratio of the fluxes strongly depends on the temperature of the star (see Fig.\ref{fig_anisotropy}). Indeed, if the IC scattering occurs deeper in the Klein-Nishina regime, then the cross section shows a weaker angular dependence. Thus, if the star temperature were  larger by a factor of 1.5, then the ratio of the fluxes would be consistent with the IC scenario, considering the relatively large error bars.

(ii) If the VHE emission production site has a size not much smaller than the distance to the optical star, then the IC scattering from each location of the pulsar will be characterized by different scattering angles, thus the orbital phase dependence of the VHE flux will be smoothed.

(iii) The effects of \g-\g~absorption, anisotropic IC scattering and Doppler boosting  may compensate each other. Thus the observed flux accidentally appears at the same level. It has to be noted, however, that as for photon absorption  this possibility requires a higher luminosity of the optical star, otherwise the \g-\g~attenuation is negligible \citep{kirk}.

While the second possibility requires a very detailed modelling of the system, which would cover self-consistently the hydrodynamics of the interaction, particle acceleration and non-thermal emission production, the points (i) and (iii) (except for Doppler boosting) may be checked through a detailed study of the properties of the optical star in the system.\\ 
X-ray observations can provide important information for the study of the leptonic scenario. The same electrons, which emit VHE emission through the IC scattering, will produce X-ray synchrotron emission. Moreover, this channel has certain advantages as compared to VHE emission: (i) X-rays are emitted isotropically, independent of the orbital location (unless the magnetic field has an ordered structure); (ii) X-rays should not suffer absorption (given the large size of the system). However, these advantages are compensated to some extent by an additional {\it a priori} unknown parameter, the magnetic field in the production region. Considering the magnetic field being only dependent on the separation distance between the two objects \citep{kirk,khangulyan}, one should expect the X-ray fluxes to be symmetric with respect to periastron passage, as is the case for  TeV \g-rays. In X-rays, however, such a behavior can not be seen. In Fig.~\ref{x-ray} X-ray data in the 1$-$10 keV band from ASCA \citep{kaspi,hirayama}, Beppo-Sax, XMM Newton \citep{XMM,suzaku}, Swift, Chandra and Suzaku \citep{uchiyama,suzaku} from the last periastron passages since 1994 are shown as a function of the binary separation $r$. Here the picture is different from the symmetric lightcurve seen in the VHE regime. For $r>2\times10^{13}\,\mbox{\rm cm}$ pre- and post periastron phases are separated by almost one order of magnitude in flux since there is a steep drop in the pre periastron part of the lightcurve (black symbols ). Interestingly, this point of broken symmetry in the X-ray lightcurve is roughly  coincident with an increase in the TeV flux by a factor of $\sim3$ (compare Fig.~\ref{curve}). The clear discrepancies between the X-ray and the \g-ray lightcurves presented here may indicate that the non-thermal particles responsible for the corresponding emission could be different or that the magnetic field has a rather complicated dependence on the orbital phase.  

\section{Description of the model}
 \label{model}
In this section, the model that was used to calculate the TeV energy flux of \psrb~taking into account adiabatic cooling of VHE electrons is briefly explained. For a more detailed description see \citet{khangulyan}.

The electron steady state distribution function (EDF) $n_{e}(t,\gamma)$ of shock accelerated electrons as a function of time $t$ (i.e.~orbital position) and energy $\gamma$ can be described by the following expression (see e.g. \citet{ginzburg}):
\begin{equation}
\label{eq:elec_dis}
n_{e}(t,\gamma) = {1\over|\dot{\gamma}(\gamma,t)|} \int\limits^\infty_\gamma Q(\gamma'){\rm d}\gamma'
\end{equation}
Here, the electron energy loss rate $\dot{\gamma}$ due to e.g. synchrotron radiation (syn), IC cooling (ic) and adiabatic losses (ad) can be written as
\begin{equation}
\label{losses}
\dot{\gamma} = \dot{\gamma}_{\rm syn}+\dot{\gamma}_{\rm ic}+\dot{\gamma}_{\rm ad}\,.
\label{eq:loss}
\end{equation}
While the calculation of $\dot{\gamma}_{\rm syn}\propto\rm B^2\gamma^2$ and $\dot{\gamma}_{\rm ic}$ is straight forward (see e.g. \citet{bosch-khang}) it is highly non trivial to compute $\dot{\gamma}_{\rm ad}$ analytically since it depends on the regime of the hydrodynamical interaction between the PW and the stellar outflow. Therefore, the adiabatic cooling profile is inferred by comparing an interpolation of the VHE data with the expected lightcurve for constant electron injection, i.e.~$\dot{\gamma}_{\rm ad}=0$ (see \citet{khangulyan} their figure\,2).

The other notations in Eq.\,(\ref{eq:elec_dis}) are the following. The source term $Q(\gamma)$ denotes the incident injection spectrum of shocked electrons from the pulsar wind. Given the lack of understanding of the acceleration processes in binary systems,  it was assumed to be a simple power law with an energy cutoff at $E_{e,max}$:
\begin{equation}
\label{e_injection}
Q(t,\gamma) = A\gamma^{-\alpha}e^{-\frac{\gamma mc^{2}}{E_{e,max}}}\theta\left(\gamma-\gamma_{\rm min}\right)\,,
\end{equation}
where $\theta$ is a Heaviside function. In general, this kind of acceleration spectrum with $\alpha\sim2$ is consistent with observations of pulsars \citep[e.g.][]{kennel_coroniti}. The spectrum is defined by the following parameters:  $A$ is the normalization coefficient, determined by the fraction of PW particles transferred to VHE injection particles; $\alpha$ is the slope of the accelerated spectrum and is defined by the acceleration mechanism, and $\gamma_{\rm min}$ is the minimum energy in the accelerated spectrum. The cutoff energy $E_{\rm e,max}$ should be determined by the balance between acceleration and losses. A simple estimate for the corresponding acceleration timescale, which is a function of the magnetic field $B(r)$, which itself depends on the binary separation $r$ as $B(r)\propto1/r$ , is
\begin{equation}
t_{\rm acc}\approx \eta\frac{R_{\rm L}}{c}\,,
\label{eq:accel}
\end{equation}
where $R_{\rm L}=mc\gamma/eB(r)$ is the Larmor radius for leptons and $\eta>1$ characterizes the acceleration rate. Comparison of Eq.\,(\ref{eq:loss}) and Eq.\,(\ref{eq:accel}) provides the value of $E_{\rm e,max}$. The EDF obtained with the inferred adiabatic loss profile was used to calculate the IC fluxes. Since adiabatic losses are assumed to be  dominant, the EDF is nearly independent of  the strength of the magnetic field. Attenuation of the TeV emission due to \g-\g~absorption appears to be non-significant for \psrb~\citep{kirk} and thus has not been considered in the calculation for the VHE fluxes presented here. 
  
\section{Results}
 \label{results}
\begin{figure}[h!]
\centering
\resizebox{\hsize}{!}{\includegraphics{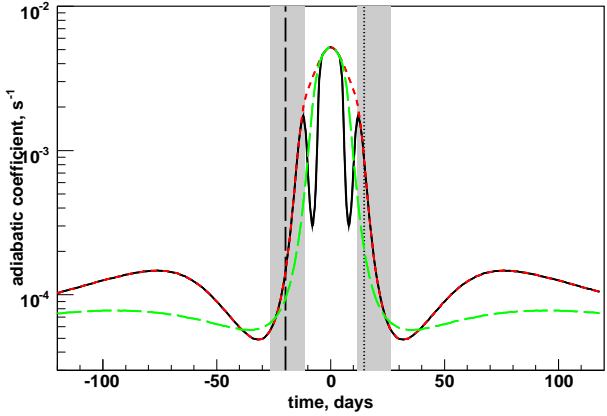}}%
\caption{Cooling coefficients for \psrb~as inferred from the combined \hess~datasets shown in Fig.~\ref{curve}. The grey bands indicate the extension of the stellar disc. The vertical dashed and dotted lines correspond to the location of the stellar equatorial plane \citep{Bmz05}. }
\label{cool}
\end{figure}

\begin{figure}
\centering
\resizebox{\hsize}{!}{\includegraphics{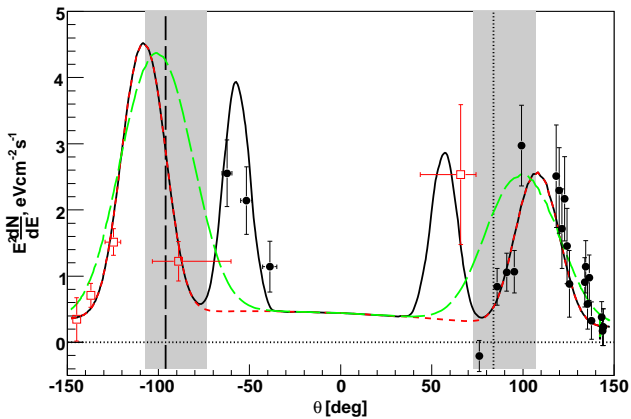}}%
\caption{Predicted energy fluxes of 1 TeV photons from \psrb~corresponding to the three different non radiative loss profiles as shown in Fig.~\ref{cool} compared to the 2004 (black dots) and 2007 (red empty squares) \hess~data. The vertical grey boxes indicate the location of the stellar disc. The vertical dashed and dotted lines denote the position of the stellar equatorial plane \citep{Bmz05}.}
\label{lc_model}
\end{figure}

In this study, three different time profiles for the non radiative cooling coefficients as shown in Fig.~\ref{cool} were adopted in order to account for the $\gamma$-ray flux variability in \psrb~as measured by \hess~ For the calculations, the following orbital parameters were used: eccentricity $e=0.87$, periastron separation $r_{\rm 0}=9.6\times 10^{12}{\rm cm}$, apastron separation $r_{\rm a}=1.4\times 10^{14}{\rm cm}$ and inclination of the orbit  $i\simeq35^\circ$ \citep{john92}. Using these non-radiative loss profiles for $\dot{\gamma}_{\rm ad}=\dot{\gamma}_{\rm ad}(t)$ in Eq.\,(\ref{losses}), the EDF $n_{\rm e}(t,\gamma)$ is obtained through Eq.\,(\ref{eq:elec_dis}). Thus, calculations of the IC energy flux using the newly computed EDFs that additionally account for adiabatic losses give the predictions shown in Fig.~\ref{lc_model}.\\
For this calculation, it was assumed that $10\,\%$ of the pulsar spindown luminosity was transferred to non-thermal particles. The magnetic field at periastron was chosen to be $B=0.1\,\rm G$. An energy range of 1 GeV to 10 TeV for the injected electron spectrum was used. Other parameters were assumed to be the same as in \citet{khangulyan}. In comparison, the latter study employed an adiabatic loss profile inferred from \hess~2004 data only, not accounting for the drop in the VHE photon flux at period $\tau\approx17$ d corresponding to $\theta\approx\pm90^{\circ}$. \citet{uchiyama}, on the other hand, assumed the time scale of adiabatic losses around $\tau\approx30$ d to be $\sim$10 times shorter than indicated in this study in order to match the broadband spectral energy distribution of the system while otherwise adopting similar parameters and radiation processes.\\
Comparing Fig.~\ref{curve} (bottom panel) and Fig.~\ref{lc_model} it seems that there is a peculiar ``dip'' in the TeV emission around $r\approx 2\times10^{13}\,\mbox{\rm cm}$ that is best reproduced when assuming a non-monotonic loss rate as depicted by the black solid curve in Fig.~\ref{cool}. Using a monotonic and thus less complicated cooling function such as the green dash dotted profile gives also a good overall agreement with the \hess~data but fails to describe the reduced photon fluxes seen at $\theta\approx\pm90^{\circ}$.\\
The predictions of X-ray fluxes (1-10 keV) for the adopted non-radiative loss profiles and model parameters are shown in Fig.~\ref{x-ray}. The pre periastron data (black symbols) deviate considerably from the model, which adopts a symmetric behavior of the $B$-field with respect to periastron. Obviously this assumption alone does not explain the X-ray data.  %

\section{Discussion \& Conclusions}
\label{concl}
\psrb~has been observed for several orbits over the past decades exhibiting a globally consistent picture for the phase dependent electromagnetic radiation at radio wavelengths, X-Rays and VHE (see e.g.~\cite{uchiyama} their Fig.~2). This suggests a physical link between the emission mechanisms in the different energy regimes such as the increasing radiation efficiency with decreasing binary separation distance $r$ resulting in enhanced PW and stellar wind interactions. However, a more detailed comparison between the VHE and X-Ray lightcurves as a function of $r$ reveals substantial qualitative differences in terms of symmetry with respect to periastron and evolution in flux level. While in the X-Ray lightcurve, Fig.~\ref{x-ray}, pre- and post periastron data are separated by almost one order of magnitude in flux level for $r>2\times10^{13}\,\mbox{\rm cm}$ such an asymmetry can not be seen in the TeV band as shown in Fig.~\ref{curve}. Even if the TeV data suffers from low statistics, a similarly  prominent feature in this energy regime should be notable. This is also reflected in the unexpected high TeV flux at $r\sim3.6\times10^{13}\,\mbox{\rm cm}$ ($\theta\sim-125$ deg) when associating the TeV emission with the disc location proposed in \citet{XMM}. The latter was inferred mainly from post periastron data (see \cite{1259new} their Fig.~7) demanding for an asymmetric lightcurve due to a tilted stellar disc with respect to the pulsar orbital semi minor axis. However, it has to be noted that at present the discrepancy of the X-ray and TeV data, in the above discussed context, relies mainly on this single pre periastron VHE point. Another deviation in the two energy bands relates to the increase in pre periastron X-ray fluxes seen for decreasing $r$ in the range $2.4\times10^{13}\,\mbox{\rm cm}>r>1.5\times10^{13}\,\mbox{\rm cm}$. This evolution appears to be reversed in the VHE lightcurve where there is a significant drop in the TeV photon flux between $r\sim2\times10^{13}\,\mbox{\rm cm}$ and $r\sim1.5\times10^{13}\,\mbox{\rm cm}$.\\
Taking into account the orbital orientation of \psrb~with respect to the observer, Doppler boosting of post shock particle flows should affect the  TeV lightcurve and yield strong asymmetries which are not seen in Fig.~\ref{curve}. Of course Doppler boosting could in principle always be accidentally compensated by other factors such as anisotropic IC. 
In general, however  (and especially in case of an IC scenario), the VHE flux from a binary system is not expected to be symmetric with respect to the periastron passage. Indeed, the corresponding symmetric orbital phases use to be characterized by different \g-\g~opacities and scattering angles with respect to the line of sight. The flux symmetry with respect to periastron passage, indicated in the case of \psrb, suggests either a rather fine compensation of the above mentioned effects or  (i) a negligibly small absorption of the \g-rays and (ii) a \g-ray production mechanism, which has a weak dependence on the orbital phase. The latter could be the case for e.g.~a large production region in \psrb~or IC scattering to proceed in the deep Klein-Nishina regime with the cross section $\sigma_{\rm ic}$ only changing marginally with $\theta$. %

To explain the variability of the observed flux as a function of the separation distance one possibility is  to introduce non-radiative (adiabatic) losses, which would dominate in \psrb~over the whole orbital period (see Fig.\,4 in \citet{khangulyan}). Moreover, the observed lightcurve with a number of humps and dips suggests a rather complicated dependence of the non-radiative losses on the separation distance. Given the complexity of the self-consistent calculation of the adiabatic losses, TeV data were used to infer a possible profile of the losses. In particular, one may expect the adiabatic loss rate to have a peak close to periastron together with two smaller peaks located at orbital positions characterized by a true anomaly of $\theta\approx\pm75^{\circ}$.  Those smaller peaks may be linked to the impact of the  stellar disc, as indicated in Fig.~\ref{cool}, when the pulsar exits the equatorial wind and thus the loss rate due to interference with the outflow decreases.\\
The predicted TeV lightcurves including different possible cooling profiles, Fig.~\ref{cool}, give an overall qualitative agreement with the data (see Fig.~\ref{lc_model}). Profiles with a simple evolution of the cooling rate such as the dashed curves in Figs.~\ref{cool} and \ref{lc_model} do not account for the TeV data close to periastron ($\theta\approx\pm70^{\circ}$). Naturally, the best agreement is achieved by a cooling function featuring two additional peaks that account for the potential impact of the stellar disc (black solid curve in Figs.~\ref{cool} and \ref{lc_model}). The predicted lightcurve shows the moderate impact of anisotropic IC scattering while still being qualitatively compatible with the observations.\\
Regarding X-ray emission, the predicted lightcurves corresponding to the presented cooling coefficients only show a weak degree of agreement to the data (Fig.~\ref{x-ray}). The most prominent disagreement stems from pre periastron data for binary separation distances $r>2\times10^{13}\,\mbox{\rm cm}$. This basically indicates a more complicated $r$-dependence of the magnetic field in the production region than the symmetric $B(r)\propto1/r$ assumed in this study. %
There is qualitative agreement for the post periastron lightcurve (red symbols) as far as flux level and global evolution is concerned. Introducing a second peak close to periastron in the black solid model curve seems also not justified in X-rays even if improving the prediction compared to the red dashed simple curve. In conclusion, none of the suggested model curves accounts quantitatively for the observational data in this energy regime based on a simple phase dependence of the $B$-field.  %
This  requires additional assumptions on the orbital phase dependence of the magnetic field.\\ 
Although any detailed discussion of the ratio of TeV and X-ray fluxes requires rather accurate calculations and goes beyond the scope of this paper, a qualitative explanation for a sharp increase of the X-ray flux before periastron passage may be suggested here. In the framework of dominant non-radiative losses, a decrease of the IC flux by a factor of 3 should be caused by an equivalent enhancement of the non-radiative losses. This may be achieved with an increase of the ram pressure in the stellar outflow, so that the PW termination shock moves closer to the pulsar (roughly by a factor of 3). Such a scenario is naturally provided by the pulsar crossing the dense stellar disc.  This would not significantly change the density of the target photons (assuming the shock is located close to the pulsar), but should lead to a significant increase of the magnetic field in the production region (again by a factor of 3). %
Thus, it is indeed natural to expect an increase of the synchrotron flux by one order of magnitude. On the other hand, it has to be noted that the overall orbital behavior of the X-ray flux cannot be explained by means of these simple arguments. In this regard it is noteworthy that the increase in pre perisatron X-ray emission coincides with the position of the stellar disc (see black points and vertical dashed line in Fig.~\ref{x-ray}). The steepest increase of the emission is roughly aligned with the equatorial stellar plane where the disc density is presumed to be highest. Again, it is difficult to account for the behavior of the post periastron data (red points) in this model picture. 

Although the \hess~data are not yet significant enough to be conclusive regarding the existence of two dips in the TeV lightcurve, the symmetry of the presented feature in the VHE regime with respect to periastron and the correlation with a significant rise of the X-ray emission at the same separation distance are certainly notable. It could be well explained by a non monotonic adiabatic cooling profile, tracing a change in the shock region's size and magnetic field conditions induced e.g.~by stellar matter outflows of increased density such as the stellar disc.

Future observations in the VHE regime with instruments such as CTA or \hess~II should shed light on this interesting question.

\begin{acknowledgements}
The author wants to thank F.~Aharonian and D.~Khangulyan for fruitful discussions and support.
\end{acknowledgements}
\bibliographystyle{aa}
\bibliography{14338}

\begin{thebibliography}{26}
\expandafter\ifx\csname natexlab\endcsname\relax\def\natexlab#1{#1}\fi

\bibitem[{{Aharonian} {et~al.}(2005)}]{1259}
{Aharonian}, F. {et~al.} 2005, A\&A, 442, 1

\bibitem[{{Aharonian} {et~al.}(2006)}]{5039}
{Aharonian}, F. {et~al.} 2006, \aap, 460, 743

\bibitem[{{Aharonian} {et~al.}(2009)}]{1259new}
{Aharonian}, F. {et~al.} 2009, \aap, 507, 389

\bibitem[{{Aharonian} \& {Atoyan}(1981)}]{aharonian}
{Aharonian}, F.~A. \& {Atoyan}, A.~M. 1981, \apss, 79, 321

\bibitem[{{Albert} {et~al.}(2009){Albert}, {Aliu}, {Anderhub}, {Antonelli},
  {et~al.}}]{LSI2}
{Albert}, J., {Aliu}, E., {Anderhub}, H., {Antonelli}, L.~A., {et~al.} 2009,
  ApJ, 693, 303

\bibitem[{{Bogomazov}(2005)}]{Bmz05}
{Bogomazov}, A.~I. 2005, Astronomy Reports, 49, 709

\bibitem[{{Bogovalov} \& {Khangoulyan}(2002)}]{bogo&khang}
{Bogovalov}, S.~V. \& {Khangoulyan}, D.~V. 2002, Astronomy Letters, 28, 373

\bibitem[{{Bogovalov} {et~al.}(2008){Bogovalov}, {Khangulyan}, {Koldoba},
  {Ustyugova}, \& {Aharonian}}]{bogovalov}
{Bogovalov}, S.~V., {Khangulyan}, D.~V., {Koldoba}, A.~V., {Ustyugova}, G.~V.,
  \& {Aharonian}, F.~A. 2008, MNRAS, 387, 63

\bibitem[{{Bosch-Ramon} \& {Khangulyan}(2009)}]{bosch-khang}
{Bosch-Ramon}, V. \& {Khangulyan}, D. 2009, International Journal of Modern
  Physics D, 18, 347

\bibitem[{{Chernyakova} {et~al.}(2009){Chernyakova}, {Neronov}, {Aharonian},
  {Uchiyama}, \& {Takahashi}}]{suzaku}
{Chernyakova}, M., {Neronov}, A., {Aharonian}, F., {Uchiyama}, Y., \&
  {Takahashi}, T. 2009, \mnras, 397, 2123

\bibitem[{{Chernyakova} {et~al.}(2006){Chernyakova}, {Neronov}, {Lutovinov},
  {Rodriguez}, \& {Johnston}}]{XMM}
{Chernyakova}, M., {Neronov}, A., {Lutovinov}, A., {Rodriguez}, J., \&
  {Johnston}, S. 2006, MNRAS, 367, 1201

\bibitem[{{Ginzburg} \& {Syrovatskii}(1964)}]{ginzburg}
{Ginzburg}, V. \& {Syrovatskii}, S. 1964, Pergamon Press, Oxford

\bibitem[{{Hirayama} {et~al.}(1996){Hirayama}, {Nagase}, {Tavani}, {Kaspi},
  {Kawai}, \& {Arons}}]{hirayama}
{Hirayama}, M., {Nagase}, F., {Tavani}, M., {et~al.} 1996, \pasj, 48, 833

\bibitem[{{Johnston} {et~al.}(2005){Johnston}, {Ball}, {Wang}, \&
  {Manchester}}]{J05}
{Johnston}, S., {Ball}, L., {Wang}, N., \& {Manchester}, R.~N. 2005, MNRAS,
  358, 1069

\bibitem[{{Johnston} {et~al.}(1992){Johnston}, {Manchester}, {Lyne}, {Bailes},
  {Kaspi}, {Qiao}, \& {D'Amico}}]{john92}
{Johnston}, S., {Manchester}, R.~N., {Lyne}, A.~G., {et~al.} 1992, ApJL, 387,
  L37

\bibitem[{{Kaspi} {et~al.}(1994){Kaspi}, {Tavani}, {Nagase}, {Hoshino}, {Aoki},
  {Hirayama}, {Kawai}, \& {Arons}}]{kaspi}
{Kaspi}, V.~M., {Tavani}, M., {Nagase}, F., {et~al.} 1994, in Bulletin of the
  American Astronomical Society, Vol.~26, Bulletin of the American Astronomical
  Society, 1485--+

\bibitem[{{Kawachi} {et~al.}(2004)}]{kawachi}
{Kawachi}, A. {et~al.} 2004, Apj, 607, 949

\bibitem[{{Kennel} \& {Coroniti}(1984)}]{kennel_coroniti}
{Kennel}, C.~F. \& {Coroniti}, F.~V. 1984, Apj, 283, 694

\bibitem[{{Khangulyan} {et~al.}(2007){Khangulyan}, {Hnatic}, {Aharonian}, \&
  {Bogovalov}}]{khangulyan}
{Khangulyan}, D., {Hnatic}, S., {Aharonian}, F., \& {Bogovalov}, S. 2007,
  MNRAS, 380, 320

\bibitem[{{Khangulyan} {et~al.}(2008){Khangulyan}, {Aharonian}, {Bogovalov},
  {Koldoba}, \& {Ustyugova}}]{khangulyan_hepro}
{Khangulyan}, D.~V., {Aharonian}, F.~A., {Bogovalov}, S.~V., {Koldoba}, A.~V.,
  \& {Ustyugova}, G.~V. 2008, International Journal of Modern Physics D, 17,
  1909

\bibitem[{{Kirk} {et~al.}(1999){Kirk}, {Ball}, \& {Skjaeraasen}}]{kirk}
{Kirk}, J.~G., {Ball}, L., \& {Skjaeraasen}, O. 1999, Astroparticle Physics,
  10, 31

\bibitem[{{Takata} \& {Taam}(2009)}]{takata}
{Takata}, J. \& {Taam}, R.~E. 2009, \apj, 702, 100

\bibitem[{{Tavani} \& {Arons}(1997)}]{tavani&arons}
{Tavani}, M. \& {Arons}, J. 1997, \apj, 477, 439

\bibitem[{{Tavani} {et~al.}(1994){Tavani}, {Arons}, \& {Kaspi}}]{Tavani94}
{Tavani}, M., {Arons}, J., \& {Kaspi}, V.~M. 1994, \apjl, 433, L37

\bibitem[{{Uchiyama} {et~al.}(2009){Uchiyama}, {Tanaka}, {Takahashi}, {Mori},
  \& {Nakazawa}}]{uchiyama}
{Uchiyama}, Y., {Tanaka}, T., {Takahashi}, T., {Mori}, K., \& {Nakazawa}, K.
  2009, ApJ

\bibitem[{{Wang} {et~al.}(2004){Wang}, {Johnston}, \& {Manchester}}]{wang}
{Wang}, N., {Johnston}, S., \& {Manchester}, R.~N. 2004, \mnras, 351, 599

\end{thebibliography}

\end{document}